 \documentclass[manuscript]{acmart}
 \usepackage{multirow}

\AtBeginDocument{%
  \providecommand\BibTeX{{%
    \normalfont B\kern-0.5em{\scshape i\kern-0.25em b}\kern-0.8em\TeX}}}

\setcopyright{acmcopyright}
\copyrightyear{2022}
\acmYear{2022}
\acmDOI{XXXXXXX.XXXXXXX}

\acmConference[Conference acronym 'XX]{Make sure to enter the correct
  conference title from your rights confirmation emai}{June 03--05,
  2018}{Woodstock, NY}
\acmPrice{15.00}
\acmISBN{978-1-4503-XXXX-X/18/06}

 \usepackage{todonotes}

\begin{document}

\title{Needs and Challenges of Personal Data Visualisations in Mobile Health Apps: User Survey}

\author{Yasmeen Anjeer Alshehhi}
\email{yanjeeralshehhi@deakin.edu.au}
\affiliation{
  \institution{Deakin University}
  \country{Australia}
}

\author{Mohamed Abdelrazek}
\email{mohamed.abdelrazek@deakin.edu.au}
\affiliation{
  \institution{Deakin University}
  \country{Australia}
}

\author{Ben Philip}
\email{benjo@deakin.edu.au}
\affiliation{
  \institution{Deakin University}
  \country{Australia}
}

\author{Alessio Bonti}
\email{a.bonti@deakin.edu.au}
\affiliation{
  \institution{Deakin University}
  \country{Australia}
}

\renewcommand{\shortauthors}{Trovato and Tobin, et al.}

\begin{abstract}
Personal data visualisations are becoming a critical contributor toward the successful adoption of mobile health (m-health) apps. Thus, understanding user needs and challenges when using mobile personal data visualisation is essential to ensuring the adoption of these apps. This paper presents the results of a user survey to understand users' demographics, tasks, needs, and challenges of using mobile personal data visualisations. We had 56 complete responses. The survey's key findings are: 1) 51\% of the users use multiple health tracking apps to achieve their goals/needs; 2) bar charts and pie charts are the most favourable charts to view health data; 3) users prefer to visualise their data using a mix of text and charts - explanation is essential. Furthermore, the top three challenges reported by the participants are: too much data displayed, overlapping text, and visualisations are not helpful in information exploration. On the other hand, users' top three encouragement factors are easy-to-read presented data, 
easy to navigate, and quality data are shown in the chart. Furthermore, fun and curiosity are the primary drivers of m-health tracking apps. Finally, based on survey results, we propose data visualisation designing and developing guidelines that should avoid the reported challenges and ensure user satisfaction.
In future work, we plan to contextualise our study and investigate the pain and gain of data visualised in the following m-health domains: sports activities, heart monitoring, blood pressure, sleeping pattern, and eating habits.
\end{abstract}



\maketitle

\section{Introduction}
Health tracking apps are one of the most growing app categories on app stores attracting developers, researchers, and investors’ interest \cite{6}. In 2016, Quintiles IMS \cite{2} published a report confirming over 165,000 health apps developed for diet and fitness tracking purposes. In addition, a recent survey indicated that 25\% of adults use mobile apps for health care purposes \cite{1}. With the anticipated increase in the number of smartphone users \cite{23}, the number of health trackers and health tracking apps is projected to increase.
Non-expert users are users with basic to no experience in data visualisation (with different gender, ages, education levels, races and socio-economic backgrounds). They are the primary target users of health tracking apps. These users rely on mobile data visualisations - \textit {e.g. charts such as bar, line, stacked bar and bullet graphs that translate their raw data and numbers into user-friendly charts} - to make decisions about their health. This places mobile data visualisation at the centre of building successful, effective and widely-adopted health tracking apps \cite{6}. Existing design thinking practices address this issue by engaging end-users in early design steps to ensure building the right apps, and data visualisation requirements \cite{8}.

With the increase number of the m-health apps, researchers conducted multiple studies to investigate the tracked data, charts used to present these data and level of users satisfactions toward the content of these apps and it UX (user experience) design. For example, Oh et al. \cite{6} reviewed 100 self-tracking tools and highlighted 3 categories of health data being tracked including: body information (height, weight, blood pressure, heart monitor, etc.), psychological state (stress, mood, emotions and well-being), and activity (sleep, food, and exercise). The authors highlighted that time series graphs are widely used to present data changes over time. They also highlighted that 1/3 of users in US have graph literacy - ability to read graph data - which makes it difficult for users to engage with the presented charts. 
Other studies investigated the charts used to present m-health data. for exmaple Katz et al. \cite{20} explored the effectiveness of data visualisation in diabetes commercial apps. They evaluated 6 different apps (Mysugar, SiDiary, iBGStar, Bant, Roche, and Diabetik) and recruited 13 participants who were diagnosed with diabetes. The participants’ technical background was in software design, graphic design, technology and diabetes. The authors outlined issues related to mobile data visualisation primarily the small screen which made it difficult for users to explore their data, and increased their cognitive load to remember data presented across a sequence of screens which led to losing context.

The authors in \cite{41} used mHealth app reviews - from Google Play - to understand user challenges and concerns related to data visualisations. The authors listed 18 data visualisation issues including: chart functionality, such as missing charts based on the entries, chart interactivity, such as zooming the chart to access more details, charts layout and style. The authors highlighted the impact of the poor data visualisations on user experience and app overall rating.

Although the app reviews analysis helped to understand user concerns better, it was mainly limited to the feedback provided by the users in their reviews without a systematic data collection process - i.e. a user study. Thus, no study captures the challenges and needs of mobile data visualisations in m-health apps.
This work complements \cite{41} with a user survey investigating users’ preferred charts, everyday tasks, and what they like and dislike about data visualisations in these m-health apps. The key research questions we aimed to address in this paper are:
\begin{itemize}
    \item RQ1: What are the top m-health apps for good data visualisations? - this question would help us recommend to developers and designers examples of top apps(the frequent apps that will be mentioned by respondents) to check when designing their data visualisations.
\item	RQ2: What are the preferred data visualisation types in terms of ease of reading and understanding? - this question would help to curate and recommend a list of visualisation charts that users can understand
\item	RQ3: What are the most common data visualisation tasks and goals users may want to achieve? - this would help designers think about what visualisation tasks to consider in their apps.
\item	RQ4: What do users like or dislike in data visualisations in these apps?
\item	RQ5: What are the key user requirements to improve their experience when using data visualisation in these apps?
\end{itemize}

The research team conducted a user survey recruiting m-health app users from different online interest groups and the public to address these research questions.
The rest of the paper is organised as follows: Section 2 presents the related work, then Section 3 presents the study design. Section 4 presents survey results and addresses the presented research questions. Section 5 presents the analysis of the identified gaps and presents recommendations for mobile data visualisations. Section 6 presents the threat to validity. Finally, Section 7 presents the conclusion and future work. 

\section{Related Work}
\subsection{Mobile visualisation}
According to Games and Joshi\cite{26}, web visualisation, mobile visualisation and dedicated visualisation environment are the three favourable environments to adopt data visualisation beyond desktop screens. However, mobile data visualisation has proposed main limitations related to its screen size and different interaction methods\cite{23}. Therefore, commendable efforts have focused on mobile device data visualisation, some of which focused on designing data visualisation interfaces. Others focused on the suitability of charts displayed on smartphone screens. For example, Roberts et al. \cite{25} highlighted that a compact data visualisation must be built to fit the screen size and interaction modality (touch and tap). Additionally, in 2018, Brehmer et al. \cite{27} investigated viewing ranges over time on mobile devices by adopting two layouts: linear and radial range charts. Both layouts possess a similar number of errors. In radial layout, the errors were in reading value and locating Max/Min. In linear layout, the errors were in locating Max/ Min value and comparing ranges of values. Despite these errors, participants performed tasks quicker in linear than in the radial chart. Therefore, this study indicates that linear and radial charts can be displayed on smartphones. In an extended study, the same authors \cite{28} conducted an additional experiment to find the efficiency of using small multiples and animation in scatter plot graphs for mobile devices. The tasks performed were comparison trends, and the output of this experiment was that both visualisations suited mobile devices when applied with large data sets.

On contrast, limited studies focused on data visualisation design for smartphone devices. For example, in 2015, Games and Joshi \cite{26} conducted a user study on Roambi (commercial visualisation system in iPad). The study recruited 24 participants aged (18-50) to evaluate the system and its visualisation. The evaluation involved providing feedback on the level of difficulty in understanding the visualised data and how well they could interact with it. The authors developed a list of recommendations for mobile data visualisations based on participants’ responses. However, these guidelines are limited to simplified data visualisation layout and interface navigation functionality aspects only. A further limitation of this study was that it approached tablet devices, which has a bigger screen than a smartphone. Thus, it is unclear if these recommendations apply to smartphone devices. In 2018, Gu, Mackin and Zheng \cite{44} presented a prototype of a data visualisation app of health data using PhoneGap - hybrid plat- form that supports the development of cross-platform mobile applications. Based on 20 responses, the authors highlighted three main concerns: colour usage, complex data visualisation interface, and lack of ideal data visualisation. Although, the efforts mentioned above have made an opening step in addressing data visualisation challenges for smaller screens, especially in the UX aspect.

Regarding mobile visualisation and Human-computer interaction (HCI), Lee et al. \cite{23} have published a paper that focuses on providing accessible mobile data visualisation for non-expert users. In 2021, Jena et al. \cite{30} published a study that provided characteristics of the non-expert users of mobile data visualisation and argued that this area needs further investigation. Thus, we argue that a great effort has been made in evaluating the suitability of some of the data visualisation types, but not the suitability of mobile data visualisation with users.

\subsection{Self-tracking and mobile health apps}

The main self tracking activities that covered in the research studies are sports activities \cite{52}, nutrition \cite{51}, health conditions \cite{33}, memories tracking \cite{34}, mood tracking \cite{35} and sleeping habits \cite{50}. As mentioned in the introduction, the number of self-trackers is projected to gradually increase due to easy access to smartphones and tracking apps. However, research studies have indicated unsolved challenges related to these mobile tracking apps.

For instance, Kyoung et al. \cite{51} conducted an experiment that aimed to understand quantified users and their interaction with tracking apps. They post this experiment in a quantified self forum and ask users to record videos and answer 3 questions: what did you do, how did you do it, and what did you learn. Based on the received 52 videos, the authors summarised their key findings: health condition was the main tracked data, and the presented data was too much, making users give up tracking and analysing their data. By answering the question "How did you do it?" participants reported the spreadsheet as the primary tool to analyse their data and commercial hardware for data collection. However, users wanted to have their own tool in which they could customise their tracking goals and charts to use. They also complained about the complexity of reading their charts due to a lack of scientific knowledge. Thus, developing simple m-health apps is needed to serve this group of people to help them analyse their personal data.

As data visualisation is the central part of the tracking data, Bongshin et al. \cite{32} published a workshop proposal to investigate mobile devices' opportunities for end-users. They identified multiple apps that have been developed for self-tracking. Examples of these apps are 1) sleep tight, which visualises sleep patterns by showing sleep duration and quality. 2) ConCap, which shows diabetes patients their data over a timeline. 3) OmniTrack included a dashboard of charts presenting different data types that the app collected. However, the authors claimed the lack of methods to evaluate data visualisation on mobile apps.

These apps take us to a new mobile health app (m-health) terminology. Significant efforts have been made in this area, focusing on providing a specific framework to develop a well-defined contextual m-health app \cite{37}, and another study focused on m-health designing based on customers' experiences \cite{38}. Furthermore, further effort is related to providing a security framework for m-health apps in data analysis and visualisation \cite{38}.

It is agreed that great efforts have been made in mobile data visualisation. However, some gaps have not been bridged yet. These gaps are related to the content and design of charts of m-health apps targeting multiple users. Additionally,  a set of best practices to evaluate these mobile data visualisations has not been studied yet \cite{32}, specifically in terms of fitting these data visualisations with small screens and the suitability of the presented data visualisation with user preferences regarding the chart types, tasks, and interactions. 

Our focus in this paper is the end-user experiences by finding their preferences in terms of explored charts and frequent tasks they applied when using m-health apps. With further investigation, we would apply users' challenges and advantages when exploring m-health data through data visualisation.


\section{study design}
We designed the study based on the design thinking process and the value-proposition canvas (Figure 1) as the two widely adopted techniques to build a successful and well-received product \cite{45}, \cite{16}. The value proposition canvas is a method that assists in ensuring a product/ service is aligned with the user’s needs and values. It has two main components:

 \begin{itemize}
\item The circle on the right is the customer profile: it focuses on identifying the customer (users), their Goals (what they try to do), their pains (challenges to achieve these jobs), and their gains (what they want to achieve) \cite{24}.
\item The square on the left is the value proposition:it has three main points to address: Gain creators, pain relievers and the final products or services. The collected survey responses and the solution to address users’ expectations should assist in delivering these three points. This will be explained further in the Survey analysis section \cite{24}.
 \end{itemize}
 
 \begin{figure}[h]
  \centering
  \includegraphics[width=\linewidth]{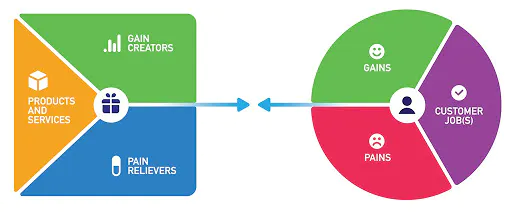}
  \caption{Value proposition canvas}
\end{figure}
\begin{figure*}[h]
  \centering
  \includegraphics[width=.95\linewidth] {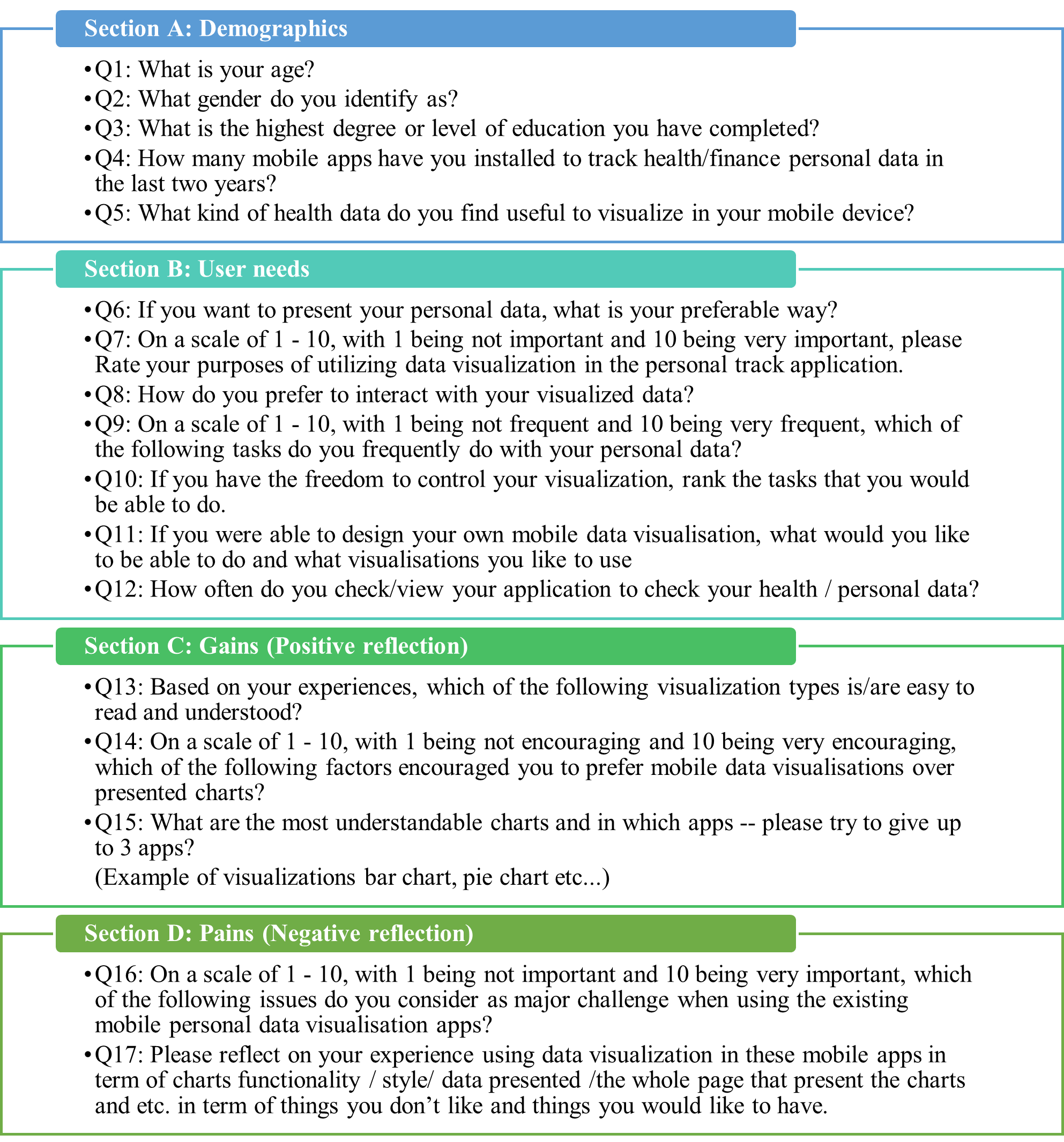}
  %
  %
  \caption{\label{fig:firstExample}
            Questions distribution based on customer profile aspects}
\end{figure*}
We built our survey based on the customer profile in the value proposition canvas. It has two parts, the introduction, which includes a brief description of the project and the questions part, in which respondents have to answer all the provided questions. The survey includes 17 questions. 15 were close-ended questions with 7 multiple-choice questions that allowed a single answer, 3 multiple choice questions that allowed multiple answers, 4 slider questions in which respondents can use the scale from 1 to 10 to answer the question, and 1 ranking question. We included only 2 open-ended questions in this survey. 

Figure 2 shows the survey questions distribution based on the customer profile (The circle in Figure 1). It displays user demographics and users’ needs. It also presents the gains aspect reflecting the positive experiences of data visualisation in m-health apps. Finally, it presents pain-related questions, reflecting users’ negative experiences using data visualisation in m- health apps. 

After we got the human ethics approval, we published the survey through social media channels, including Facebook, Instagram, Reddit and Twitter. In addition, we published an anonymous link and QR code in survey flyer to ensure reaching multi-users interested in tracking their personal data. The survey data is stored anonymously. 

\section{Survey results}
In total, 180 participants accepted the consent form, and out of these, only 56 completed the entire survey. We checked the completed responses to ensure these were not bots by reviewing the written responses, and all 56 responses had actual inputs. In the following sub-sections, we present these responses’ results based on the 56 completed responses and findings.

\subsection{Participants Demographics}
In this section, we report on the demographics of the participants. Age: 28.57\% of the participants were (18-30), 35.71\% of the participants were (31-40) years old, 21.43\%
were (41-50) and 14.29\% of participants were above (50) years old. Gender: 55.36\% of participants were females, 42.86\% were males and 1.79\% have selected prefer not say . Education level: 50\% of the participants had Bachelor’s degrees, 30.36\% earned a master’s degree, 3.75\% completed their post-secondary and upper education, and 12.50\% earned a PhD degree. Number of m-Health apps used per person: 41.07\% have used only one app to track their health data, 39.29\% have installed fewer than five apps, 12.50\% have installed more than five apps, and few participants have not found a suitable app yet 7.14\%. These demographic data reflect a reasonable and diverse range of participant’s profiles in the survey.

\subsection{RQ1: Top Health Apps in terms of good data visualisations}
In order to answer this research question, we asked the participants to submit their top three apps in terms of the most understandable charts. Table 1 shows a list of all health apps that participants have reported. 5 apps have been mentioned frequently in the collected responses. Apple health app, Sweatcoin, Samsung health, fitness apps and step tracker pedometer.
\begin{table}
  \caption{list of health apps}
  \label{tab:freq}
  \begin{tabular}{cccl}
    \toprule
     Health apps & No. of downloads & User ratings\\\\
    \midrule
   Apple Watch apps & Built-in &3.3 \\

 Apple health app& Built-in & 4.3\\
 
 Samsung health & 1,319,723 & 4.0\\
 
 Fitness apps & 485,134 & 4.0\\
 
 Sweatcoin & 362,465 & 4.5\\
 
 Running apps & Users have not specify the app& --- \\
 
 Step app &52,808 & 4.6 \\
 
 HealthifyMe & 1,000+ & 4.6 \\
 
 Zepp &   555,777& 4.4 \\  
 
Huawei health&  622,971 &4.4 \\
 
 MySejahtera& 812,011 & 4.9\\
 
 Period tracker& 6,271,138& 4.9  \\
  
Samsung health & 1,000,000+ & 3.3\\
 
 pacer & 833,72 & 4.7\\
 
 Step tracker & 433,121 & 4.6 \\
 
my care plan & 50,000+& 3.0\\
  \bottomrule
\end{tabular}
\end{table}

As health app marketing is growing continuously, we could not put a fixed list of the top health apps in the stores. However, we noticed that 4 of the mentioned apps in the responses are always shown as top health apps in the Google Play Store. i.e. HealthifyMe, Pedometer, Samsung health apps and Sweatcoin. We also found that Pedometer, Period Tracker and Apple Health Tracker are shown as top apps in App Store which also have been reported by users in the survey. 

\subsection{RQ2: Preferred data visualisation types}
Generally, 75\% of the respondents preferred to represent their data using reports including both text and charts. However, 19.64\% of the respondents have selected charts were only but it is still favourable over texts only (5.3\%). We also offered an open-ended question to collect the top charts (in terms of most understandable, accessible and easy to read). The participants reported: bar charts 23 times, pie charts 14 times, and line charts 5 times as the preferable charts to use in m-health apps. The other charts have recorded single occurrences such as star chart, table, map, timeline, and speedometer. In line with the found results, the author in \cite{51} stated that bar charts and line charts have also been shown as the most used charts among 21 other visualisation charts. Additionally, authors in \cite{20} have stated that the pie chart scored high rates for being easy to read for participants.
 \begin{figure}[htb]
  \centering
  \includegraphics[width=.95\linewidth] {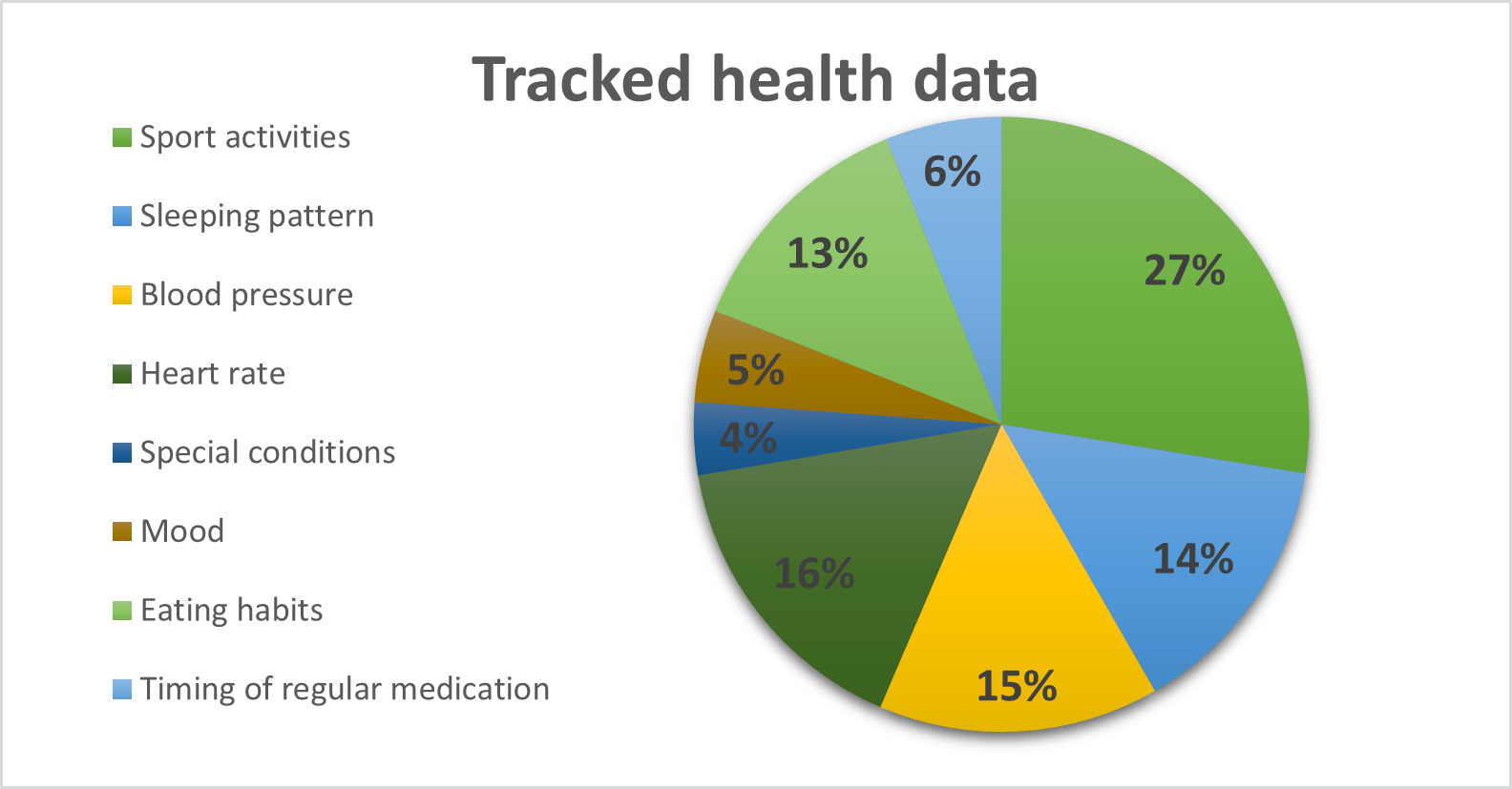}
  %
  %
  \caption{\label{fig:firstExample}
           Tracked Health Data}
\end{figure}
\subsection{RQ3: Data visualisation goals and tasks}
\textbf{Goals of tracking:} 
We provided eight health activities and asked respondents to rank them according to their importance (Figure 3). Sports activities were on the top (26.79\%). On the other hand, there were slight differences between heart rate (15.84\%), blood pressure (14.29\%), sleeping pattern (13,69\%) and eating habits (12.50\%). The minor goals of health data were timing on regular medication (5.95\%) and mood (4.76\%). This could be because of the age of 

\textbf{Critical tasks of interest:} We introduced two questions in our survey to understand the most frequent visualisation tasks and interactions users usually do. Additionally, we raised one question that helped us understand users' needs regarding tasks and interactions they wish to apply to m-health charts. 

The first question was to identify the frequent tasks that users usually do. Showing the progress of activity was the most selected task with a mean (7.21 out of max 10), finding maximum and minimum came next (6.82 out of max 10), and comparison (6.48 out of max 10  was the third option as shown in Table 2. The other tasks: filtering, ordering and showing a trend, details on demand, identifying patterns and relationships have had almost similar mean scores of 5.36, 5.51, 5.41, 5.67,5.73, respectively.

The second question was to select the preferable way to interact with the visualised data. Four options were provided, however, there was a slight variance in the top three selected interaction methods. Read-only (32.14\%), read and edit visualisation (26.79\%), drill down to show details (21.43\%) and drag and drop (12.50\%). These percentages indicate that users still prefer to read their visualisation with minimum interaction and not interested in complex visualisations. 

The third question was to rank the tasks users would like to have when exploring their charts. We provided 11 options and asked respondents to rate them from 1 to 10. To have a detailed view of users' wanted to have tasks, we sum up 1,2,3,4,5 ranking to be the minimum and 6,7,8,9,10 to be the maximum (minimum vs maximum).
Editing chart title (17 vs 27) was the top task users would like to do. Next, show the history of the previous visualised data (22 vs 26) was the second. Then, finding a pattern in the data (17 vs 26) was the third. After that, sorting data (27 vs 23), cluster data (26 vs 23), changing colours (22 vs 23), and comparison tasks (27 vs 22) have had the same popularity. Finally, touch on the point to get details (28 vs 20). Rank data (28 vs 20), View charts (28 vs 21) and changing values to change the results in the charts (35 vs 14) were the least preferred tasks. The total of these numbers may vary as some users have not rated some of these tasks.
\begin{table}
\caption{Top 3 preferred tasks}
\label{tab:freq}
  \begin{tabular}{ccl}
    \toprule
 Top Tasks & mean\\
\midrule
Showing the progress of activity  & 7.21\\
Finding maximum and minimum & 6.82 \\
Comparison & 6.48 \\
  \bottomrule
\end{tabular}
\end{table}
\subsection{RQ4: Issues and encouragement factors of mobile data visualisation} 
\subsubsection{Top issues that affect m-health data visualisation: }
We framed two questions to identify the issues and challenges of m-health data visualisation from users' points of view. The first was a text entry question in which users could type their problems \textit{(this was at the beginning of the survey to avoid leading answers)}. To analyse the open-ended questions, each member of the research team mapped the answers to data visualisation components (interaction, data , functional requirement and style). Then we reviewed our mapping to validate it and agreed on mapping strategy.
Participants gave different answers related to their dislikes of the presented visualised data in their apps. As figure 4 shows, most participants have issues with the data presented (14 answers), and only two participants have written interaction as an issue. Furthermore, users reported issues in charts (8 answers) and style (5 answers). On the other hand, the most needed features that participants reported were charts (14 answers) and interaction (8 answers), and the least were style (7 answers) and data (5 answers). Unfortunately, users had not detailed these issues. Thus, it is hard to acknowledge what was the concern about charts, interactions, data presented and style.
We proposed a detailed question to mitigate the broad answers and avoid misleading answers in the above question. Thus, we asked respondents to rank a given 27 issues on a scale from 1 to 10, with 1 being not important and 10 being very important (or concerning). \textbf{These issues were sourced from our previous work from m-health app reviews \cite {41} and the systematic literature review work \cite{40}. }
\begin{figure}[htb]
  \centering
  \includegraphics[width=.95\linewidth] {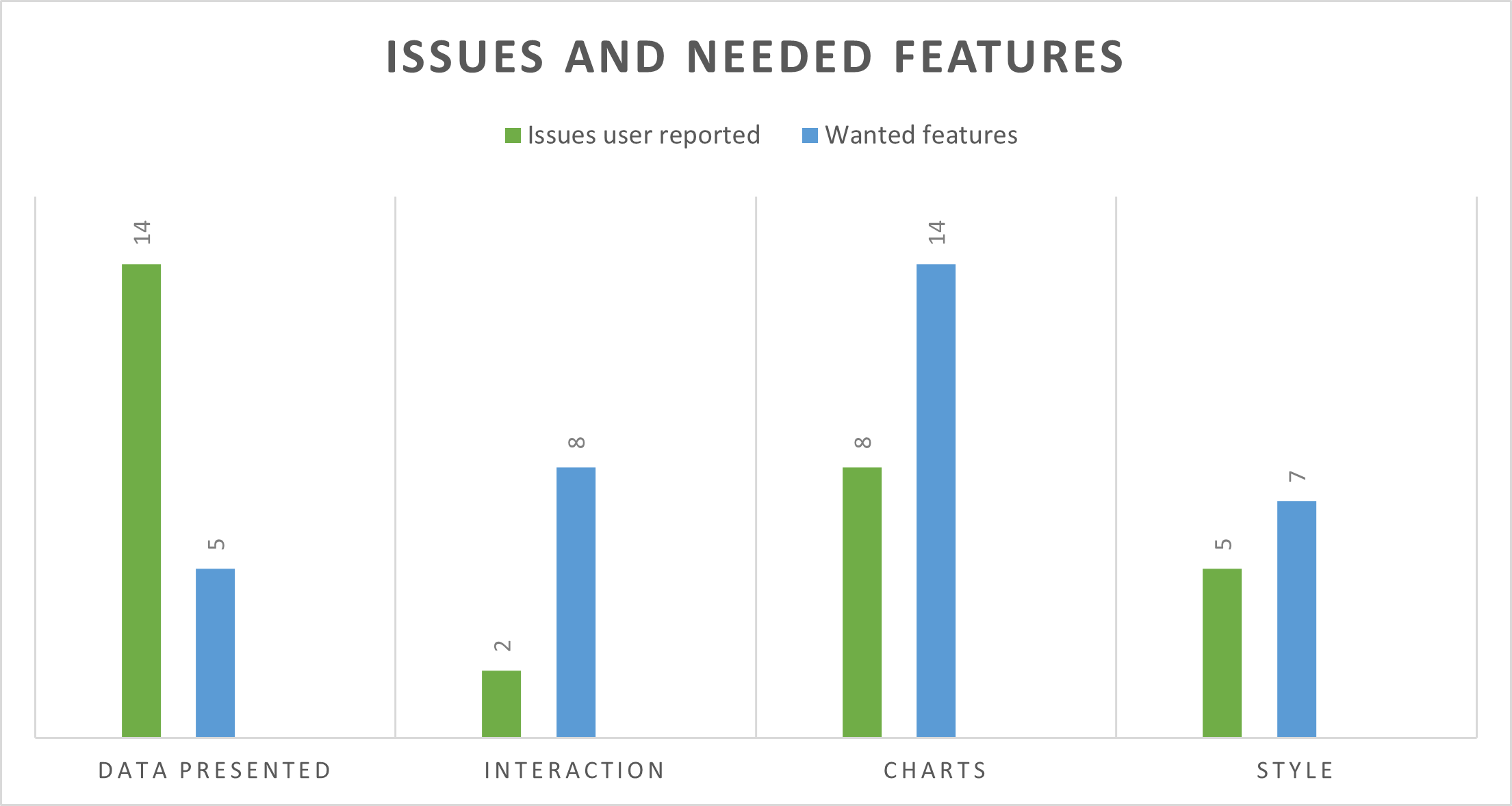}
  %
  %
  \caption{\label{fig:firstExample}
           Issues and needed features.}
\end{figure}

Table 3 shows the list of the top 10 challenges users reported when using data visualisations. We extracted these top issues based on the sum of rating from 6 to 10. These issues covered multiple aspects related to developing and designing data visualisation - e.g. charts functionality, displayed data, chart style and interaction. 
The challenges related to the displayed data aspect were the highest-rated challenges. - e.g.Data displayed is too much, overlapping text, visualisation is not helpful in information exploration, and charts are not showing the needed information. Charts interaction challenges were the second top-rated issue. Users rated scrolling up and down to compare two charts. The touch interface is not precise, difficulty dragging points and a lack of control over data visualisation as fourth, fifth, seventh and ninth issues, respectively. The third aspect that users were concerned about was chart functionality. They rated I cannot see the chart of my entries as the sixth concerned challenge. Finally, in regards to chart styling, users rated titles are not clear as the eighth challenge.

On the other hand, the second part of Table 3 shows the less concerned issues based on users' ratings. These issues are related to chart style, device interactivity and displayed data. The challenges related to chart style are the most rated issues. -E.g. 1) legends are missing, 2) lack of tool-tip, 3) presented data is not suitable with backgrounds, 4) colours are not presenting meaningful information, 5) charts are not suitable in terms of data presented, and 6) grid scaling is not clear. Regarding device interactivity, users rated zoom in and out are not applied as the fifth less concerning issue. Users rated that they cannot derive insights from the presented visualisation as a minor issue regarding the displayed data aspect. Finally, the lack of consistency is a less concerning issue that affects the entire m-health app.
\begin{table}

\caption{Top and bottom 10 data visualisation challenges with sum of ratings}
\label{tab:freq}
  \begin{tabular}{ccl}
    \toprule
 Top Challenges & \# responses ranked +6 \\ [0.1ex] 
 \midrule
 Data Displayed is too much & 45 \\ 

Overlapping text & 40  \\
 
Visualisation is not helpful in exploring information & 39 \\
 
 Scrolling  up  and  down  to compare two charts & 38\\
 
 The touch interface is not precise & 38\\  
 
 I can not see the chart of my entries & 37\\ 
 
Difficulty to drag points & 36 \\
 
 Titles are not clear & 34 \\
 
Lack of control over data visualisation & 33\\
Charts is not showing the needed information & 32\\
\end{tabular}
  \begin{tabular}{cc}
     \toprule 
       Bottom Challenges & \# responses ranked -6 \\ 
 \midrule
 Can not derive insights out of the presented visualisation & 24 \\
 
 Legends are missing & 24 \\
 
 charts are not suitable in terms of data presented & 24  \\
 Zoom in and out is not applied & 24 \\
 
 I can not understand the presented chart & 24\\
 
 Lack of tool tip & 23 \\

 Grid scaling is not clear & 23 \\
 
 Presented data is not suitable with the background & 21  \\

 Colours are not presenting meaningful information & 19 \\ 
 Lack of consistency & 19 \\ 
 \bottomrule
  \end{tabular}
  
\end{table}

\subsubsection{Top factors that affect acceptance of data visualisation:} We provided 19 statements in which users rated them from 1 to 10 where 1 indicated the less encouragement and 10 indicated the most encouragement. Table 4 shows the top 10 encouragement factors based on the mean statistics. Interactivity has four factors: 1) easy to navigate, 2)ability to set goals, 3) easy to explore and compare between visualised data 4) Interaction with the visualised data is accessible. Three of these factors are related to displaying data which are easy to read presented data, data are complete and shown in the chart, and data presented are correct. Functionality aspect had got one statement which is chart show progress. One statement related to device compatibility: the chart is fitted with a mobile screen. Finally, one statement related to chart styling is that iconography helps track progress. 

We analysed the collected response based on (the minimum rated vs the maximum rated). All of the statements related to interactivity had got high rates which are:  Ability to set your own goal (9 vs 43), ease to explore and compare between visualised data, ease to navigate (11 vs 42), having control over the presented charts (15 vs 38). The chart presented can be re-scaled using zooming in/out (16 vs 37), and interaction with the visualised data is accessible (16 vs 37). Displayed data statements were the second most top-rated, which are: data are complete and shown in the chart (7 vs 44), easy to read presented data (6 vs 48), and data presented are correct (7 vs 46). Chart style has got favourable rates: colours make visualisation easy to read, font type and size are clear (8 vs 45),  visualisation charts convey the needed information - titles are understandable. Iconography helps track progress (11vs 42). The device compatibility statement had a positive rating: charts presented with the mobile screen were the sixth (11 vs 42). The lowest encouragement factors were related to the design aspect: consistency of design (14 vs 39). Finally, The three encouragement factors that scored the worst were related to functional requirements complete set of charts that represent all entries (13 vs 33), Clutter-free (23 vs 31) and tooltip is helpful (20 vs 33).

\begin{table}
\caption{Top 10 data visualisation encouragement factors}
\label{tab:freq}
  \begin{tabular}{ccl}
    \toprule
  Factor &  Mean\\ [0.1ex] 
 \midrule
 
Easy to read presented data & 8.20 \\
 
 Easy to navigate & 7.93 \\

Data are complete and shown in the chart & 7.63 \\
 
 Charts show progress & 7.56\\
 
 Ability to set your own goal & 7.54\\  
 
Data presented are correct & 7.48\\
 
Easy to explore and compare between visualised data &7.28\\
 
The chart presented is fitted with a mobile screen size &7.28 \\
  
Iconography is helpful in tracking progress & 7.06\\
Interaction with the visualised data is easy & 7.00 \\
\bottomrule
\end{tabular}
\end{table}

\subsection{RQ5: Suggestions to improve data visualisation}

To accumulate user requirements for better data visualisation design, we proposed an open-ended question in which participants had two choices: write their suggestions or upload a sketch illustrating their thoughts on how to improve visualisation. Most participants (86.96\%) agreed to write their suggestions rather than upload sketches. As this is an open-ended question, each research team analysed and filtered the responses based on their validity - i.e. responses were written in English and meaningful. Then, we sat together to review each researcher's analysis and make a unified analysis of each validated response. The responses covered multiple aspects of data visualisation components. Therefore, we grouped these suggestions into five covered aspects. 
\textbf{Audiences:} age-friendly data visualisation is one of the reported recommendations.Users reported that “some data visualisation are not age-friendly in font size and colours. ``Making data visualisation user friendly by providing straightforward access'' was another suggestion that multiple users frequently reported. 
\textbf{Functional requirements:} users suggested multiple functional requirements in which there were requests on: ``add face scan to know the mood'', ``Being able to compare between current and history data'', and ``need for the daily dashboard in which user can personalise their presented data''. Moreover, although tool-tips were not one of the top-rated challenges, one end user reported a need for a user guide that does not consume the memory and battery of the smartphone.
\textbf{Data:} ``I think one of the most important things is for charts not to be crowded, and as a non-expert, I need to be able to understand the two parameters being compared easily``, ``describing data more understandably''. 
\textbf{Interactivity:} one user reported a suggestion on having control over the charts ``I would like to add high and low limit lines to the Y-axis on a graph so that any Y values above the high limit are easily seen as above or below those lines''. Additionally, users are looking for more interactive health trackers apps. They suggested to add fun and attractive features such as adding emojis. Finally, users suggested to add interactive charts that represent their health data.
\textbf{Summary:} As mobile health apps are used by diverse users of different ages, educational levels, and cultures, this leads to diverse preferences. Such as preferred data visualisations, the purpose of health tracking, tasks they are trying to achieve, encouragement and challenging factors in using charts. By investigating the top-rated challenges that are shown in Table 3, it is clear that our findings align with \cite{20} where the researchers examined users’ experience in using commercial diabetes apps. They found that users complained about the limitations of interactivity, inadequate data presentation, screen size limitations, and colour selection. Furthermore, although study \cite{26} included suggestions for better chart layout and simply navigation guidelines, these two aspects are still proposing issues based on user perspectives. Hence, the current mobile data visualisation design and development practices must be revised and improved to address these challenges.
\section{SURVEY ANALYSIS}
The left side of the value proposition in Figure 1 focuses on a value map that consists of product and service, pain relievers, and gains creator \cite{24}. In our case, the service is to support designers in delivering well-designed and developed m-health apps based on the data visualisation perspectives. Ideally, there should be a framework or guidelines to help visualisation designers during this process. We have identified two main industrial guidelines developed for data visualisation design: \textit {Google material design} \cite{47} (BETA version that would be changed based on research studies) and \textit{IBM design language} \cite{46}. Figure 5 shows the covered aspects in these two industrial guidelines.

\begin{figure}[htb]
  \centering
  \includegraphics[width=.95\linewidth] {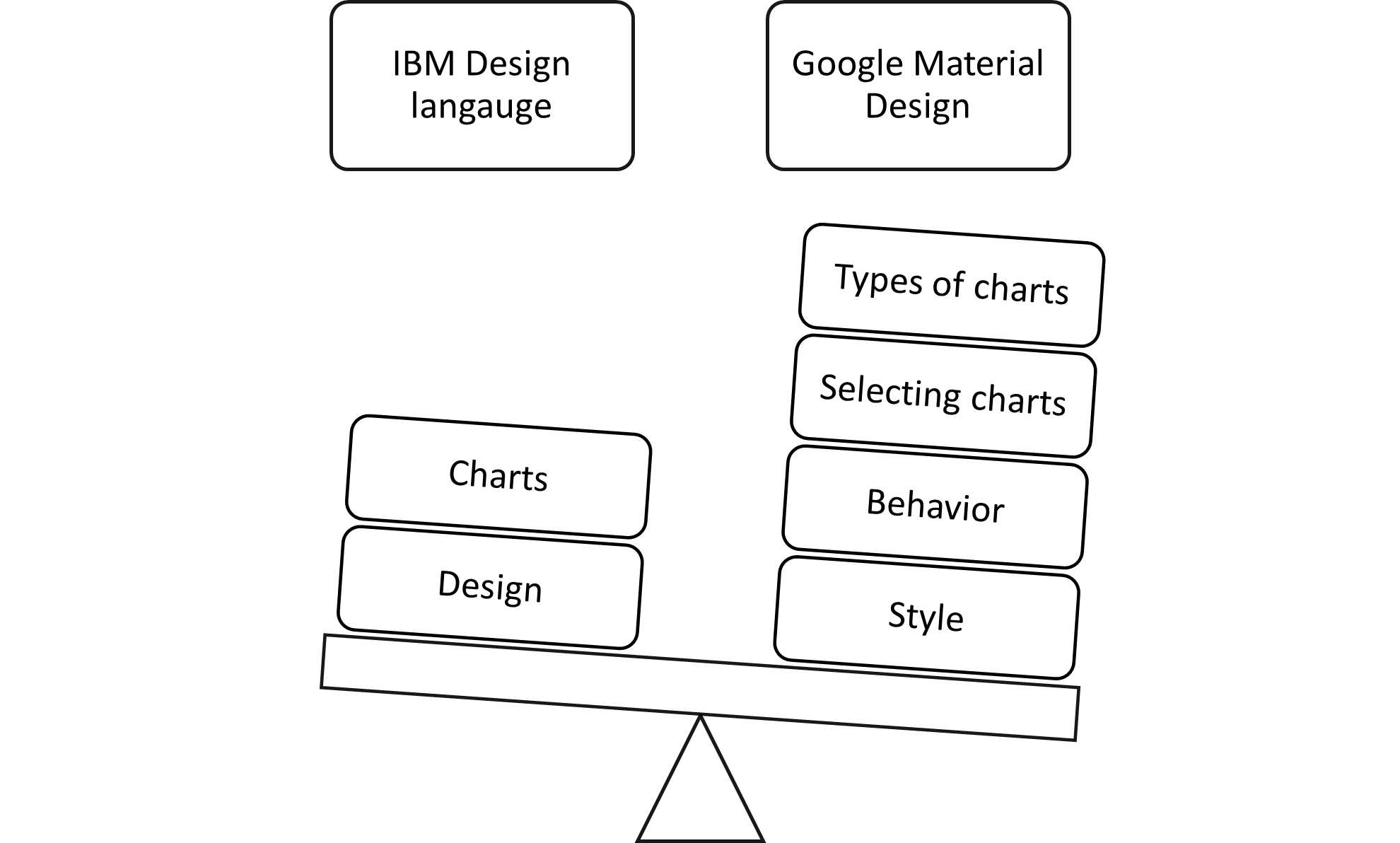}
  %
  %
  \caption{\label{fig:firstExample}
           Covered aspects in IBM and Google Design language}
\end{figure}

There are 3 similar aspects addressed by both Google and IBM guidelines (chart selection, chart design, chart behaviour):
\begin{itemize}
    \item Chart selection: both have explained the usage of different charts. So, it guides designers in selecting the right chart to present the data they want to communicate.
    \item Chart design: both design languages provided chart shape and colour guidelines.
    \item Google has provided another section called behaviour. The main interactions were getting details of the presented points and zooming interaction considering the mobile device environment. On the other hand, IBM has mentioned interactions in the design section, which enable users to go from chart overview to the more detailed by using the filtering function.
\end{itemize}
The guidelines provided in Google material design are more comprehensive than the IBM design language. Google material design considered chart interactivity based on the mobile device platform. Its design language supported touch, tap interaction, pagination by swiping to the left to view the other charts, panning and zooming using 2 fingers to pinch. However, these guidelines suffer from limitations of other components related to data visualisation, such as considering non-expert audiences, presented data, chart requirements, chart interactivity and alignment of presented chart design and functionality with the used mobile devices. Thus, they are not appropriate, and hence there is a need for a revised framework. Below, we outline a set of essential requirements (dimensions) for such guidelines based on our survey results and user reviews and comments reported in \cite{40}
\begin{itemize}
    \item Target audience: the guidelines need to address users’ domain terms of the level of experiences, culture, age, health condition and socio-economic. This would ensure a complete, correct and consistent list of targeted users.
    \item 	Functional requirements: the guidelines need to ensure a complete list of functional requirements that are related to the health tracking domain.
    \item Data: the guidelines need to maintain the completeness, correctness and consistency of data presented in the charts.
    \item 	Looks and feels: this aspect has been addressed well in the frameworks above. However, the guidelines must ensure that styling, colouring and chart design are correctly aligned with the targeted audiences.
    \item 	Interactivity: although this aspect has gotten some attention in Google material design. However, the guidelines must ensure that these interactions are completely based on users’ needs. Moreover, the guidelines should ensure correct and consistent chart interactivity.
    \item Mobile platform: the guidelines need to ensure that the presented chart and its interactivity are suitable for the user’s mobile device
    \item 	Single/App visualisation: the guidelines need to ensure that the data visualisation in the m-health app covers all the aspects mentioned above.
\end{itemize}

Concerning gain creator and based on the survey results, we created a list of positive points that should be considered when designing m-health data visualisation. The order of this list is based on the relevance to the nice-to-have scale.
\begin {itemize}
\item {Bar charts, pie charts and line charts are the most favourable charts, the designer should consider using these charts whenever it is possible }
\item {Designers should also consider the top encouragement factors that are listed in Table 4 }
\item {Designers should consider these apps: Apple health app, Sweatcoin, Samsung health, fitness apps and step tracker pedometer as users have frequently mentioned them in their responses}
\end {itemize}

Finally, regarding the pain relievers, we also created a list of how to overcome the highly concerning challenges. 
\begin {itemize}
\item {Present only the needed data}
\item {Ensure the style of the charts is appropriate to overcome overlapping text}
\item {Examine the provided charts and their content before publishing them}
\item {Provide a suitable interaction with the charts and avoid intensive scrolling }
\item {Consider the device in use, and it is interaction}
\item {Make sure all functional requirements are working well}
\item {Ensure the look and feel of the charts are well designed and enable an option of controlling the font size, type and positing }
\end {itemize}

The outcome of the gains creators and pain relievers can be combined and make a set of guidelines that help designers build data visualisation for m-health apps. 

\section {THREATS TO VALIDITY}
 \textbf{Internal validity:} Although we built our survey based on the value proposition canvas and following the thinking process, we assume that we may miss some of the challenges, encouragement factors, and users’ needs in our survey. The reason behind that is that targeting non-expert users is not easy and needs a careful understanding of the users’ domain. We also acknowledge the missing of some demographic factors, such as socioeconomic and culture, as we are targeting public end-users. We have piloted the survey internally within the team to capture as many details related to participants and data visualisations as possible. 
 
 \textbf{External validity:} we are targeting non-expert end-users of m-health apps. Therefore, we expected more than 100 complete responses. However, we received only 56 total responses, which may affect the diversity of the preferred options and raise challenges. In addition, other factors would affect the diversity in the responses. These factors are:
\begin{itemize}
    \item The educational background, as 50\% of the respondents are bachelors, which may affect the non-educated users.
    \item Age of the participants, out of the 56 responses, we got only 8 responses from users above 50 years.
    \end{itemize}
However, we targeted diverse social networks and interest groups to ensure we get quality responses - participants have experience using health tracking apps.

\section {CONCLUSION and future work}
This paper introduced a survey to investigate user needs, challenges, and goals for data visualisations in m-health apps. The study was designed based on the well-known value proposition canvas with three components covering user tasks, pains and gains. We have collected 56 complete responses. Our key findings include that: bar and pie charts are favourable charts. Fun, curiosity and tracking progress are the primary purposes of using charts. These charts need natural language descriptions to make them clear to multiple users. Easy to read, easy to navigate, and the completed data in the charts are the top three encouragement factors that make users frequently use their m- health tracking apps. On the other hand, too much-displayed data, overlapping text and visualisations are not helpful in information exploration were the top challenges users encountered when using m-health apps. In order to cope with these challenges and ensure user variances, we suggested that mobile data visualisation need attention in term of developing and designing aspect.
This paper presented the user perspectives about m-health apps in general. In addition, this investigation provides two directions of extension and future work.
\textit{The first is related to studying m-health domains specifically:} we plan to extend our work by contextualising each m- health domain and studying its specific challenges. We will consider the most rated health domains in the survey: sports activity, heart monitoring; blood pressure; sleeping pattern, and eating habits.
\textit{The second is related to the gap found in the design and the provided guidelines:} we plan to study the provided academic studies related to data visualisation and industrial approaches to bridge the gap in data visualisation design for non-expert users.

\bibliographystyle{ACM-Reference-Format}
\bibliography{sample-base}
\end{document}